\documentclass[aps]{revtex4}

\def\be {\begin{equation}}
\def\ee {\end{equation}}
\def\ba {\begin{eqnarray}}
\def\ea {\end{eqnarray}}
\def\nn {\nonumber}
\def\a  {\alpha}
\def\b  {\beta}
\def\c  {\gamma}

\def\d  {\delta}
\def\D  {\Delta}
\def\e  {\epsilon}

\def\k  {\kappa}
\def\l  {\lambda}
\def\m  {\mu}
\def\n  {\nu}

\def\O  {\Omega}
\def\p  {\pi}

\def\r  {\rho}

\def\s {\sigma}

\def\la {\label}
\def\le {\left}
\def\ri {\right}
\def\pa {\partial}
\def\f {\frac}

\def\no {\noindent}
\def\bi {\begin{itemize}}
\def\ei {\end{itemize}}

\def\ra {\rangle}
\def\vs {\vspace}

\def\bc {\begin{center}}
\def\ec {\end{center}}

\begin{document}

\title{Black Hole Thermodynamics: Entropy, Information and Beyond
\footnote{Plenary talk given at the 
{\it Fifth International Conference on Gravitation and Cosmology},
Cochin, 7 January 2004. } }

\author{Saurya Das}
\affiliation{Department of Physics, The University of Lethbridge, \\
4401 University Drive, Lethbridge, Alberta, CANADA T1K 3M4 \\
Email: saurya.das@uleth.ca}
\keywords{black hole thermodynamics, quantum gravity}
\pacs{04.60.-m 
04.70.-s 
}

\begin{abstract}
We review some recent advances in black hole thermodynamics, 
including statistical mechanical origins of black hole entropy 
and its leading order corrections, from the viewpoints of various 
quantum gravity theories. 
We then examine the information loss problem and some possible 
approaches to its resolution. Finally, we study some proposed
experiments which may be able to provide experimental signatures of
black holes.
\end{abstract}

\maketitle

\begin{center}

{\it {\bf Dedicated to the memory of Professor Shyamal Sengupta}}

\end{center}

\section{Introduction}

Existence of black holes are one of the most intriguing predictions
of general relativity. They are expected to have (Bekenstein-Hawking)
entropy and radiate at their 
characteristic Hawking temperature \cite{bek}. Furthermore,
these quantities should satisfy laws analogous to the laws
of thermodynamics. 
For example, for a Reissner-Nordstr\"om (RN) black hole of mass $M$ and 
charge $Q$ in $d$-spacetime dimensions with the metric:
\ba
ds^2_{RN} &=& - 
\le( 1- \f{16\p G_d M}{(d-2) c^2 \O_{d-2} r^{d-3}} 
+ \f{16\p G_d Q^2}{(d-2) (d-3) c^4 r^{2(d-3)}}  \ri) c^2 dt^2 \nn \\
&+& \le( 1- \f{16\p G_d M}{(d-2) c^2 \O_{d-2} r^{d-3}} 
+ \f{16\p G_d Q^2}{(d-2) (d-3) c^4 r^{2(d-3)}}  \ri)^{-1} dr^2 
+ r^2 d\O_{d-2}^2 ~~, 
\ea
the (outer) horizon radius, electrostatic potential at the horizon,
Hawking temperature and entropy are given by: 
\ba
r_+^{d-3} 
&=& \f{8\p G_d M}{(d-2) c^2 \O_{d-2}} 
+ \sqrt{ \le( \f{8\p G_d M}{ (d-2) c^2 \O_{d-2}} \ri)^2 
- \f{2G_d~Q^2}{ (d-2)(d-3)~c^4} } \\ 
\Phi &=& \sqrt{\f{2(d-3)}{d-2}}~\f{Q}{r^{d-3}} \\
T_H &=& \f{(d-3) \hbar c}{2\p r_+^{d-2}} 
\sqrt{\le( \f{8\p G_d M}{(d-2) c^2 
\O_{d-2}}\ri)^2 - \f{2G_d Q^2}{(d-2)(d-3)c^4}} \\
S_{BH} &=& \f{A_H}{4\l_{Pl}^2} = \f{\O_{d-2} r_+^{d-2}}{4 \l_{Pl}^{d-2}} ~ \\
\ea
where $A_H$ is the black hole horizon area. These
satisfy the zeroth, first and second laws 
of {\it Black Hole Thermodynamics}. 
\ba
T_H &=&~\mbox{constant over~~horizon} \\
d(Mc^2) &=& T_H dS_{BH} + \Phi dQ  \\  
\D S_{BH} &\geq& 0~,   
\ea
$G_d$ and $\l_{Pl}$ being the $d$-dimensional Newton's constant and
Planck length respectively, and
$\O_{d-2}$ the area of $S^{d-2}$.

One of the foremost problems in quantum gravity is to explain the origin
of Bekenstein-Hawking entropy. In other words, to discover a set of 
fundamental degrees of freedom which give rise to a (large) degeneracy 
$\O$, such that:
\be
S_{BH} = \ln\O
\ee
(where the Boltzmann constant has been set to unity). Various approaches
to quantum gravity have attempted to answer this question, with different
degrees of success. The related issue of Hawking radiation 
has also been examined in these approaches. In the following sections, 
we will review a few of these important approaches, including loop quantum
gravity and string theory. Associated with black holes is the so-called 
problem of {\it Information Loss}, which we examine in section 
\ref{infoloss} 
Finally, we review some future experiments which could shed light
on the nature of quantum gravity and test the correctness of some of the
theories. 

\section{Statistical Mechanical Origins of Entropy}

\subsection{Horizon Conformal Field Theory}

We review \cite{carlip1} and start with the Einstein action in $d$-dimensions:
\be
I = \f{c^3}{16\p G_d} \int {R} \sqrt{-g}~d^dx ~. 
\la{act1}
\ee
If one restricts oneself to the spherically symmetric sector 
with the metric:
\be
ds^2 = g_{\m\n}(x) dx^\m dx^\n + \phi(x) d\O_{d-2}^2~,~~\m,\n=t,x~~,
\ee
(where $x$ is a radial coordinate and $\phi(x) = r(x)^2/\l_{Pl}^2$)
then the 
action (\ref{act1}) 
reduces to a two dimensional dilaton gravity action of the form
\cite{gabor1}:
\be
I = \int {\cal L} \sqrt{-\c}~d^2x = \f{c^3}{2G_2} \int \le( \phi R_2 + 
\f{1}{\l_{Pl}^2} V[\phi] \ri) \sqrt{-\c}~d^2 x ~, 
\ee
where the potential $V[\phi]$ depends on the matter content of the theory.
Next, define a quantity analogous to the expansion of nulll congruences:
\be
\Theta = \f{1}{\phi} \ell^a \nabla_a \phi \equiv \f{s}{\phi}~, 
\ee
where $\ell^a$ is the null normal. Black hole horizons are characterised by
vanishing $\Theta$. Now, it can be shown that under the set of transformations:
\ba
\d g_{ab} &=& \nabla_c (f \ell^c) g_{ab} \la{tr1}  \\
\d\phi &=& (\ell^c \nabla_c h + \kappa h)~,  \la{tr2}
\ea
where $f$ is an arbitrary function, $\k(=2\p T_H/\hbar)$ the surface gravity 
and $h=sf/\k$, the variation of the Lagrangian takes the following form :
\be
{\d \cal L} \sim \Theta~, 
\ee
which vanishes at the horizon. 
Transformations (\ref{tr1}-\ref{tr2}) thus constitute an 
{\it asymptotic symmetry}. It is generated by the Hamiltonian:
\be
L[f] = - \f{c^3}{2G_2} \int_{\Delta} 
\le( 2 \ell^a \nabla_a s - \kappa s\ri) f \sqrt{\c} d^2 x~~,~
s= \ell^a\nabla_a \phi ~, 
\ee
where $\D$ is an element of the horizon.
The function $f$ can be expressed in terms of the basis functions:
\be
f_n = \f{\phi_+}{2\p s} z^n~,~z=\exp(2\p i\phi/\phi_+)~
~\mbox{with}
~ \le\{ f_m, f_n \ri\} = i(m-n) f_{m+n}  
\ee
as
\be
f = \sum c_n f_n~. 
\ee
Computation of Poisson brackets yields:
\be
\le\{  L[f_m], L[f_n] \ri\} =
{- \f{24\p s}{G_2 \k} }~\f{n^3}{12}~\d_{m+n,0} ~,
\ee
where the RHS can be identified with the central charge of the 
Virasoro algebra:
${\cal C}=-24\p s/G_2 \k$. The eigenvalue of $L_0$, $\D$ is given by:
\be
L[f_n] = {-\f{\k \phi_+^2}{4\p G_2 s}}~\d_{n0}  \equiv
\D~\d_{n0}~. 
\ee
The asymptotic density of states is then given by the Cardy formula,
whose logarithm gives the microcanonical entropy:
\be
S = \ln \rho (\D) = 2\p \sqrt{\f{{\cal C}\D}{6}}  = \f{2\p \phi_+}{G_2} 
= \f{A_H}{4\l_{Pl}^2} = S_{BH} 
\ee
This is observed to agree with the Bekenstein-Hawking entropy of the
black hole. In other words, in this approach, 
the conformal field theoretic (CFT) degrees of freedom appear to 
be responsible for black hole entropy.

\subsection{Loop Quantum Gravity}

Next, we examine loop quantum gravity (LQG) \cite{ash1}. 
Imposition of the null condition on $\D$, as well as those of no radiation
falling in and no rotation are equivalent to the condition:
\be  
\f{A_H}{2\p\c} {F}_{ab}^{AB} + {\Sigma}_{ab}^{AB} = 0
\la{loop1}
\ee
where $F$ is the field strength corresponding to the $SL(2,C)$ connection
$A_{aA}^B$ , $\s_a^{AA'}$ is the soldering form for $SL(2,C)$ spinors,
the metric $g_{ab} = \s^{AA'}_a \s_{bAA'}$, 
$\Sigma_{ab}^{AB} = 2\s^{AA'}_{[a} \s^B_{b]A'}$ ($a,b,\dots (A,B,\dots)$ 
are spacetime (internal) indices) and pullbacks of $F$ and $\Sigma$ to 
the horizon two sphere are understood.  
The gravity action can be written in terms of these variables as:
\be
I = -\f{i}{8\p G} \int Tr \le( \Sigma \wedge F \ri) 
- \f{i}{8\p G_4} \f{A_H}{4\p} 
{\int_\D Tr \le( A \wedge dA + \f{2}{3} A 
\wedge A \wedge A \ri)} 
\ee
The second term is the appropriate boundary term on $\D$, which is nothing
but the Chern-Simons action.
The quantum version of (\ref{loop1}) can be written as:
\be
\le( I \otimes \f{A_H}{2\p\c} F_{ab}\cdot r 
+ \Sigma_{ab}^{AB} \cdot r \otimes I \ri) 
{\Psi_V} \otimes {\Psi_S} = 0
\la{loop1a}
\ee
where $r$ is an internal vector and 
$\Psi_V$ and $\Psi_S$ are volume and surface states, which are
eigenvalues of the first and second terms of (\ref{loop1a}) respectively. 
It is the surface degrees of freedom which are responsible for entropy.
These puncture $\D$ in a finite number of points $n$, at each of which there
is a spin associated. Thus, the collection of surface states can be 
collectively written as:
\be
{\cal P} = \le\{ (p_1,j_{p_1}), \dots ,  (p_n,j_{p_n} ) \ri\}~, 
\ee
where $j_{p_i}$ is the spin labelling the puncture $p_i$. 

The horizon area can be thought of as being built up of individual bits
of area associated with the punctures, 
where a spin $j_p$ contributes a quantum of 
$8\p \c \l_{Pl}^2 \sqrt{j_p(j_p+1)}$, $\c$ being the
unknown Immirzi parameter. Thus:
\be
A_H = 8\p \c \l_{Pl}^2 \sum_p \sqrt{j_p( j_p +1)} ~.
\la{loop2}
\ee
(applications of LQG with a different regularisation, as well as
in the context 
of black hole coherent states, an equispaced area spectrum is
predicted, although
the area proportionality of entropy remains unchanged
\cite{alek,adg}). 
The dimensionality of the Hilbert space is a product of those for each spin, 
which is:
\be
dim\le( {\cal H} \ri) \sim \prod_{j_p \in {\cal P}} 
(2 j_p + 1) \approx 2^P ~,
\la{loop3}
\ee
where the last step follows from the fact that $j=1/2$ dominates the 
spin configuration. The number of punctures $P$ can be estimated from
(\ref{loop2}) under the same assumption, which when plugged into
(\ref{loop3}) gives us the microcanonical entropy:
\ba
S_{BH} &=& \ln dim\le( {\cal H} \ri) \nn \\
&\approx& P \ln 2 \nn \\ 
&=& \le( \f{\c_0}{\c} \ri) \f{A_H}{4\l_{Pl}^2}
~~~~,~\c_0 = \f{\ln 2}{\p\sqrt{3}} ~.
\ea
We see that the area proprtionality of entropy emerges naturally, 
although the prefactor is $1/4$ only for 
$\c=\c_0$. It may be mentioned that the value of $\c_0$ does not depend
on the specific black hole that one considers and is the same for 
dilatonic and charged black holes.

\subsection{String Theory}

Next, we turn to string theory, one of the best explored approaches 
of quantum gravity. One of the first black holes explored within 
string theory was a four dimensional charged black hole with a singular
horizon \cite{sen}. However, the one for which Bekenstein-Hawking entropy
is best explained in terms of string states is the 
five-dimensional extremal RN black hole.
This will be
reviewed in the next sub-section \cite{vs}, following which 
Anti-de-Sitter-Schwarzschild black holes will also be examined in the
context of AdS-CFT correspondence \cite{adscft}.

\subsubsection{Extremal charged black holes}

We start with the ten-dimensional low
energy effective action of Type II string theory in the strong coupling
(large $G_{10}$, equivalently large string coupling $g$) limit: 
\be
I_{} = \f{c^3}{16\p G_{10}} \int d^{10} x \sqrt{-g_{10}}
\le[  R + \f{1}{2} (\nabla {\phi})^2 - 
\f{1}{12}e^{\phi}{H_{(3)}}^2 \ri] ~,
\ee
where $\phi$ is the dilaton and $H_{(3)}$ is the $RR-3-$form field strength. 
Compactifying on a $T^4 \times S^1$ (note that originally the compact manifold 
$K3 \times S^1$ was considered), the above has a metric solution of the form: 
\be
ds_{10}^2 = {e^{2\chi}}{dx_i dx^i} 
+ {e^{2\psi}} 
({dx_5} + A_\mu dx^\mu)^2 
+ {e^{-2(4\chi+\psi)/3}} {ds_5^2}~.
\ee
In the above, $\chi$ and $\psi$ are scalar fields, $A_\m$ is a gauge
field, the first and second
terms represent metrics on $T^4$ and $S^1$ respectively, while
$ds_5^2$ is the five-dimensional extremal RN metric. 
The above configuration has
a description in terms of $D$-branes and Kaluza-Klein (KK) momenta in the
weak coupling (small $G_{10}$) limit. More precisely, the latter 
consists of $N_1$ $D-1$-branes (which couple to
$H_{(3)}$) , $N_5$ $D-5-$branes (which couple to
$^\star H_{(3)}$ ) and $N$-units of KK momenta on the $S^1$. 
Extremality condition for black holes translates to the condition
of $BPS$ saturation for these branes. In the case in which
these three charges are equal, they are related to the black hole
charge $Q$ and horizon radius $r_+$ as:
\be
N_1 = N_5 = N =
\f{Q}{\sqrt{c\hbar \l_{Pl}}} = \le( \f{r_+}{\l_{Pl}} \ri)^2~.
\la{string2}
\ee
Now, since open strings begin and end on $D$-branes, there will
be a total of $N_1N_5$ oriented strings stretching between the various $1$ and
$5$-branes (it can be shown that those that begin and end on the same brane do
not contribute to entropy to leading order). Each such string has $4$-bosonic
and $4$-fermionic degrees of freedom associated with it, corresponding to the 
four transverse directions of $T^4 \times R$ (i.e. total of
$n_B=n_F=4N_1 N_5$). Each such degree of freedom has
an energy of $N\hbar c/L$, where $L$ is the length of the $S^1$, Moreover 
$L$ is 
taken to be much larger compared to the length dimensions of the $T^4$. Then, 
the entropy of this one-dimensional gas of bosons and fermions is given 
by \cite{vs}:  
\be
S = \sqrt{\f{\p(2n_B + n_F)LE}{6\hbar c} } = 2\p \sqrt{N_1 N_5 N}~,  
\ee
which using (\ref{string2}) yields:
\be
S =  \f{\O_{3} r_+^3}{4\l_{Pl}^3} = \f{A_5}{4 \l_{Pl}^3} = S_{BH}~.
\ee
Thus we see that the entropy of the one dimensional gas {\it exactly}
reproduces the Bekenstein-Hawking entropy of the corresponding black hole. 
Furthermore, since the counting is done for BPS branes, 
supersymmetry ensures that it suffers no renormalisations
and the result continues to hold even at strong coupling.

\subsubsection{Asymptotically anti-de Sitter black holes}

Next, we consider black holes in the context of the $AdS/CFT$ correspondence. 
The metric
of an $AdS-$Schwarzschild ($AdS-SC$) black hole in $d$-dimensions is given by:
\ba
ds_d^2 &=& - 
\le( 1 - \f{16\p G_dM}{(d-2) \O_{d-2} c^2 r^{d-3}} + \f{r^2}{\ell^2} \ri)dt^2
\nn \\
&+& \le( 1 - \f{16\p G_dM}{(d-2) \O_{d-2} c^2 r^{d-3}} + 
\f{r^2}{\ell^2} \ri)^{-1}dr^2
+  r^2 d\O_{d-2}^2 
\ea
whose entropy and Hawking temperature are:
\ba
S_{BH} &=& \f{\O_{d-2} r_+^{d-3}}{4\l_{Pl}^{d-2}}
\approx c_1' T_H^{d-2}~,~~\le[~c_1'= \f{\O_{d-2}}{4\l_{Pl}^{d-2} }
\le( \f{4\p \ell^2}{\hbar c (d-1)} \ri) ^{d-2} \ri] \la{ads3} \\
 T_H &=& \hbar c \f{(d-1) r_+^2 + (d-3) \ell^2}{4\p \ell^2 r_+ } 
\approx \hbar c\f{(d-1) r_+}{4\p \ell^2},~~~r_+\gg \ell~, 
\la{ads3a} 
\ea
where  the approximation 
$r_+ \gg \ell$ is known as the `high-temperature limit'.

First we 
examine whether the 
above thermodynamic properties can be modelled by a 
`dual gas' consisting of a perfect fluid of bosons and fermions, 
residing in ${\cal D}$ 
spacetime dimensions, in some appropriate boundary of the 
black hole spacetime (Note: here ${\cal D}$ is 
identical to the $\D$ of reference \cite{husain}). We assume a 
general dispersion relation between the energy and momentum of the 
constituents of the gas, given by: $\e= \k p^\a$. ,
It can be shown that the free energy and
entropy of the gas in this case is related to 
its temperature as:
\ba
F_{gas} 
&=& - \f{c_2' V_{{\cal D}-1}}{({\cal D}-1)/\a +1 } 
T^{\f{{\cal D} -1}{\a} + 1}~, \la{free1}\\
S_{gas} &=& c_2' V_{{\cal D} -1} T^{\f{{\cal D}-1}{\a}}~, 
\la{ads2}
\ea
where 

$$
c_2'= \f{\O_{{\cal D} -2}~\le( ({\cal D}-1)/\a+1\ri)
~\zeta\le(\f{{\cal D} -1}{\a}+1\ri)~\Gamma\le( \f{{\cal D}-1}{\a}+1 \ri)
\le(n_B + n_F - \f{n_F}{2^{\f{{\cal D} -1}{\a} }}\ri) }
{({\cal D} -1) \k^{\f{{\cal D} -1}{\a}} (2\p\hbar)^{{\cal D}-1}  } ~. 
$$
%
Further, if we assume that the gas is at a distance
$r_0$, where it is in equilibrium with the Hawking radiation,
then its temperature is related to the red-shifted Hawking temperature as:
\be
 T = \f{T_H}{\sqrt{-g_{00}}} = \f{\ell~T_H}{r_0}~.
\ee
Plugging this into (\ref{ads2}), we get:
\be
 S_{gas} 
= c_2' V_{{\cal D} -1} \le( \f{\ell}{r_0} \ri)^{\f{{\cal D}-1}{\a}}~
T_H^{\f{{\cal D}-1}{\a}} ~.
\la{ads4}
\ee
Matching powers and coefficients of $T_H$ in (\ref{ads3}) and (\ref{ads4}),
we arrive at the following \cite{husain}: 
\ba
{\cal D} &=& \a (d-2) + 1
\la{ads5} \\
c_1' &=& c_2'~\f{(d-1) \O_{{\cal D}-1} \ell^{d-2} r_0^{(\a-1)(d-2)}}
{({\cal D}-1)/\a + 1} ~.
\la{ads6}
\ea
The above relations can be thought of as necessary conditions that any 
quantum theory of gravity must satisfy, if it wishes to describe $AdS-SC$ 
black holes holographically. Note that only in the case $\a=1$ 
(relativistic dispersion) is the usual holographic dimension 
(${\cal D}=d-1$) recovered.
Simultaneously, $r_0$ vanishes from (\ref{ads6}) (i.e. 
the precise location of the dual gas becomes irrelevant), suggesting perhaps that
$\a=1$ is preferred. 
The full significance of ${\cal D} \neq d-1$ is general, including fractional
${\cal D}$ is yet to be understood. 

With the above formalism at hand, let us test the validity of relations
(\ref{ads5}) and (\ref{ads6}) in the light of $AdS_5/CFT_4$ correspondence.
The dual of the black hole in this case has been conjectured to be 
${\cal N}=4$, $SU(N)$ super Yang-Mills theory for large $N$. The number
of Bosonic/Fermionic degrees of freedom and the relation between
$\ell, \l_{Pl}$ and $N$ are in this case (here $\a=1$) \cite{adscft}:
\be
N_B=N_F=8N^2~,~~ \le( \f{\ell}{\l_{Pl}}\ri)^3= \f{2N^2}{\p} ~.
\ee
Using the above, it is easy to show that \cite{adscft2}: 
\be
c_1' = \f{3}{4}~c_2'~~,~~S_{BH} = \f34 S_{gas}~.
\ee
That is, the entropies of the quite different physical systems (black hole
and gas) are almost identical! 
It has been conjectured that the discrepancy of the factor of $3/4$ is 
due to the strong vs weak coupling of the two systems, 
although a rigorous proof is lacking.

\section{Leading Order Corrections to Entropy} 

We now ask the following question: does the matching of equilibrium entropy
of black hole and its dual (as in the case of $AdS/CFT$) guarantee matching
of corrections to entropy due to thermodynamic fluctuations, which are 
always present for a thermodynamic system? In other words, we
would like to explore whether the approximate agreement becomes better or worse
when sub-leading terms are taken into account. To this end, we first 
compute the first order corretions for an arbitrary thermodynamic system, 
using the canonical framework
(for corrections due to fluctuations of geometry,
see \cite{kaulmaju}).
 . The partition function for such a system is:
\be
Z(\b)
= \int_0^\infty \r(E)~e^{-\b E} dE~,
\label{part1}
\ee
where the density of states, $\r(E)$ can be written as an inverse Laplace
transform of the partition function: 
\be
\r(E) = \f{1}{2\p i} \int_{-i\infty}^{i\infty} 
{Z(\b) e^{\b E}} d\b
= \f{1}{2\p i} \int_{c-i\infty}^{c+i\infty} e^{S(\b)}d\b~.
\la{density1}
\ee
where we have used: 
\be
S=\ln Z + \b E 
\la{ent5}
\ee
Close to the equilibrium temperature inverse $\b=\b_0~,$ 
one can expand the entropy function as:
$$ S (\b) = S_0 + \f{1}{2} (\b - \b_0) ^2 S_0'' + \cdots
\label{ent1}
$$
where
$S_0 := S(\beta_0)$ and $S''_0 := (\pa^2 S(\b)/\pa \b^2)_{\b=\b_0}$.
Substituting the above in (\ref{density1}), we get:
\be
\r(E) = \f{e^{S_0}}{2\p i} \int_{c-i\infty}^{c+i\infty}
e^{1/2 (\b - \b_0)^2 S_0''} d\b ~~~~(\b-\b_0=ix)~.
\ee
Defining $\b-\b_0=ix$ and performing a contour integration, we get:
\be
\r(E) = \f{e^{S_0}}{\sqrt{2\p S_0''}} ~~.
\la{corr0}
\ee
whose logarithm gives the corrected entropy, taking into account the 
thermal fluctuations: 
\be
{\cal S} := \ln \r(E) = S_0 - \f{1}{2} \ln S_0'' + ~\mbox{(smaller terms)}.
\la{ent7}
\ee
Next, using 
\be
E \equiv <E> = - \le( \f{\pa \ln Z}{\pa \b}\ri)_{\b=\b_0} 
= - \f{1}{Z} \le(  \f{\pa Z}{\pa \b} \ri)_{\b=\b_0}~,
~ <E^2> = \f{1}{Z} \le( \f{\pa^2 Z}{\pa \b^2} \ri)_{\b=\b_0}~  
\ee
and the definition of specific heat: 
\be
C \equiv \le( \f{\pa E}{\pa T}\ri)_{T_0}~ 
= \f{1}{T^2}  \le[ \f{1}{Z} \le( \f{\pa^2 Z}{\pa\b^2} \ri)_{\b=\b_0}  - 
\f{1}{Z^2} \le( \f{\pa Z}{\pa \b} \ri)^2_{\b=\b_0} \ri] = \f{S_0''}{T^2} 
\ee
it follows that:
\be
S_0'' = <E^2> - <E>^2 = C T^2 ~. 
\la{ent8}
\ee
This shows that it is indeed the fluctuation of the total energy 
which gives rise to corrections. 
Then from (\ref{ent7}) it follows that \cite{dmb}:  
\be
~{\cal S } = S_0 ~-~ \f{1}{2} \ln \le( CT^2  \ri) + \cdots ~.
\la{entmaster}
\ee
The above formula applies to all stable thermodynamic systems 
(for some refinements, see \cite{tran}). However, for  
black holes, we will make the substitutions 
$ S_0 \rightarrow S_{BH} = {A}/{4\l_{Pl}^{d-2}} $ and $ T \rightarrow T_H$.
For example, for a $BTZ$ black hole, 
\be
C = T_H = S_{BH} ~,
\ee
implying:
\be
{\cal S} = S_{BH} - \f{3}{2} \ln S_{BH}~. 
\ee
For $AdS-SC$ black hole one uses (\ref{ads3}), and 
(\ref{ads3a}) and the specific heat:
\ba
C &=& (d-2)\le[ \f{(d-1)r_+^2/\ell^2 + (d-3)}{(d-1)r_+^2/\ell^2-(d-3)} \ri] S_0 
\nn\\
&\approx& (d-2)~{S_0}~~~[~~\ell \ll {r_+}  ~] \la{ads9}~,
\ea
using which, we get:
\ba
{\cal S} &=& 
S_{BH} - \f{1}{2} \ln \le( {S_{BH} S_{BH}^{2/(d-2)}} \ri)  \nn\\
&=& S_{BH} - \f{d}{2(d-2)} \ln \le( {S_{BH}}{} \ri)  
\la{ads11}
\ea
For the dual gas on the other hand, from $C=d(F_{gas}+TS)/dT$
and Eq.(\ref{free1}), we get: 
\be
{\cal S} = {S_{gas}} -{\f{d}{2(d-2)} 
\ln S_{gas}}  
~~+~\f{1}{d-2} { \ln \le[
{(n_B +  n_F) V_{\D-1}} \ri]}
\ee
Note that although the second term in the RHS of the above {\it exactly}
matches the leading order correction term for the black hole
in Eq.(\ref{ads11}), there is no such counterpart of the third term. 
In other words, leading order entropy matching does not 
guarantee the matching of corrections!  This is in spite of using the
same master formula (\ref{entmaster}) 
to compute corrections, since the free energies
(and hence partition functions) of the two systems are in fact not equal. 
More work is required for a clearer understanding of this apparent discrepancy.

\section{Hawking Radiation}

Another important feature of quantum black holes is Hawking radiation (HR). 
Both LQG and ST attempt to explain this phenomenon in terms of 
microscopic degrees of freedom that are relevant for each theory. In LQG, the 
picture of HR is as follows: the states that puncture the horizon two-sphere
can jump from a higher to a lower spin state (lowering 
the horizon area), emitting a quantum of radiation in the process. 
A calculation of the corresponding radiation 
rate shows that this can indeed account for the qualitative
behaviour of HR \cite{krasnov}.

In string theory on the other hand, $D$-brane 
interactions provide the necessary mechanism.
Open strings on $D$-branes interact to form closed
strings, which leaves the brane and propagates to asymptopia. 
A careful computation of the rate in this case reproduces the HR rate 
for the $d=5$ RN black hole including all prefactors \cite{cm,dm}. 
The agreement appears to persist even in the high energy tail
of the spectrum \cite{dds}. For HR from the $AdS/CFT$ perspective, 
see \cite{ddg}.

Another explanation for HR which has been proposed does not
depend on the specific underlying theory that one is considering.
Starting from the assumption that the spectrum of the black hole horizon is 
discrete, and that the behaviour of the latter is analogous to that of an
excited atom, a discrete HR spectrum with a Planckian envelope results
\cite{mukhanov,barvin}.

\section{Information Loss from a Black Hole}
\la{infoloss}

Since a black hole emits uncorrelated thermal radiation, 
if it evaporates completely,
then the large amount of information that went inside it while it was 
being formed, is destroyed forever. This is the origin of the 
information loss problem for a black hole, which can also be understood
in the following way. Consider a Cauchy surface which intersects 
collapsing matter forming a black hole,
as well as the subsequent Hawking radiation. Now, as the black hole evaporates,
since the information that went inside the black hole cannot
be transmitted along the Cauchy surface to a region containing the observer 
ar ${\cal I}^+$, therefore the information is lost. One resolution of
this apparent paradox (since quantum mechanics at the fundamental level is 
expected to be unitary, which forbids such loss of information) 
is to assume
that the information inside the black hole 
gets `cloned' at the horizon such that
copies of every bit that is inside are transmitted via Hawking radiation to the
outside observer. Unfortunately, such cloning is forbidden by unitary quantum 
mechanics, as the following line of reasoning demonstrates. Let $|\Psi\ra$ be 
an arbitrary state to be cloned to a `target' state $|T\ra$. One can
think of the former as the page of a document and the latter as a blank  
sheet of paper in a quantum photocopying machine. Assume there is an unitary 
operator ${\cal U}$ which does the cloning, i.e. takes 
$|T\ra \rightarrow |\Psi\ra$: 
\ba
{\cal U} |\Psi\ra \otimes {|T\ra} &=& |\Psi \ra \otimes |\Psi\ra
~~,~~~{\cal U}^\dagger = {\cal U}^{-1}
\la{clone1}
\ea
The machine should also be able to photocopy another arbitrary state $|\Phi\ra$:  
\ba
{\cal U} |\Phi\ra \otimes |T\ra &=& |\Phi \ra \otimes |\Phi\ra
\la{clone2}
\ea
Taking the inner product of (\ref{clone1}) and (\ref{clone2}): 
\ba
\langle \Psi | \otimes {\langle T | {{\cal U}^\dagger~{\cal U}} 
|T \ra } \otimes |\Phi\ra &=& 
\le( \langle \Psi | \otimes \langle \Psi |  \ri)
\cdot \le( | \Phi \ra \otimes |\Phi\ra  \ri)
\\
\Rightarrow~~\langle \Psi  |\Phi\ra &=& 
\le( \langle \Psi | \Phi\ra \ri)^2 ~.
\ea
The last equation implies:

\be
\langle \Psi|\Phi \ra = 1 
~~or ~~0 ~~
\nn
\ee
meaning $|\Psi\ra$ and $|\Phi\ra$ cannot be an arbitrary quantum state. 
In other words, there is no quantum photocopier, and unitary cloning is 
impossible! 
If the above possibility is
ruled out, are there other ways of resolving this paradox?
There has been a host of other proposals, of which we mention a few here.
For other proposals, see \cite{infoother}.

\subsection{$\cal U$ is Not-Unitary} 

An obvious way to circumvent 
the non-cloning theorem is to abandon the requirement
of unitarity. If ${\cal U}^\dagger {\cal U} \neq {\cal I}$, then  
of course the above theorem does not hold and cloning is still possible 
\cite{haw2}. Although viable, this proposal is at odds with quantum mechanics
as we know it as well as most of the proposed theories of quantum gravity, 
including LQG and string theory. 

\subsection{Planck-Size Remnant}

Another conjecture put forward by various authors is that the black hole
does not radiate completely and that the evaporation stops when it reaches
Planck size. This `remnant' could then contain all the information that 
went into the black hole. While a remnant ground state is suggested in
many cases, it is far from clear whether such a large
information can be concentrated in such a small volume and more work needs
to be done in this direction \cite{remnant,barvin,adler1}. 

\subsection{Black Hole Complementarity}  

Another interesting proposal takes into account the prediction that 
even if Hawking radiation carries some information, 
an outside observer has to wait for at least half the lifetime of a black
hole to get just one bit of this information \cite{page}. 
In other words, bulk information
appears at very late times, when the black hole is almost Planck sized. 
To check whether this information obtained is indeed authentic, the 
observer can jump into the black hole, only to find that semi-classical
physics has broken down, rendering its predictions null and void.
Observers outside and inside the black hole are thus mutually 
exclusive and complementary to each other, and the above proposal is 
know as {\it Black Hole Complementarity} \cite{suss2}. Although 
this attempts to avoid the information puzzle altogether, 
at best it seems to be an effective theory, a more precise
microscopic picture being clearly warranted.

\subsection{Unique BH Final State} 

Recently, another proposal has been put forward, where it is conjectured that
all information is transmitted to outgoing Hawking radiation by a mechanism
similar to quantum teleportation, and that 
the final state of the black hole is in fact {\it unique} \cite{hormal}.  
The entropy of the final state is thus $\ln 1 =0$ and no information is lost.
However, criticisms of this proposal include the observation that entangling
interactions between the collapsing body and outgoing Hawking radiation 
spoil this unitarity even if in a weak sense \cite{gottesman}.

\section{Observations and Experiments}

The theoretical prediction of black holes is almost as old as the theory
of General Relativity itself \cite{karl}. Their fascinating properties
have been extensively studied by all quantum gravity
theories. However actual observations of black holes 
involve enormous technical difficulties. 
Here we describe some recent advances in this direction. 
For other potential signatures of quantum gravitational effects,
see \cite{ssg,smolin,amelino,jacobson,sudarsky}.

\subsection{Astrophysical Black Holes} 

There has been 
important progress in the detection of astrophysical black holes, and we 
draw the reader's attention to a review article \cite{narayan} which    
summarises compelling experimental evidences for candidate
black holes $X$-ray binaries and in galactic nucleii. 
Although these are massive, having negligible Hawking temperature, one cannot
help but speculate whether any indirect evidence for black hole thermodynamics
can be extracted by these or future observations.

\subsection{Brane World Black Holes}

There have been two recent proposals known as ADD and RS (collectively as
{\it Brane World Scenarios}) which attempt to solve the gauge hierarchy 
problem and which start with the assumption that our observed $4$-dimensional
universe is embedded in a $d$-dimensional world (the `brane') 
\cite{add,rs}. In the 
ADD scenario, which assumes that the unobserved part is a $T^{d-4}$, with 
each circle being of length $L$. the Planck
scales in lower and higher dimensions are related as:
\be
~~{M_{Pl(d)}}^{d-2} = \f{\hbar^{d-3}}{c^{d-5} G_d}~~
= \f{\hbar^{d-3}}{c^{d-5} {V_{d-4}} G_4}
= \le( \f{\hbar}{c {L} } \ri)^{d-4} {M_{Pl(4)}^2} 
\ee
From the above it follows that that even though the observed Planck energy is
$M_{Pl(4)}c^2 \approx 10^{19}-$GeV, for ~$d \geq 4$ and $L \approx 1$~mm,
the fundamental Planck
energy $M_{Pl(d)}c^2$ could be as low as $1-TeV$. In other words, the hierarchy
problem does not exist in the full spacetime. Now, energies of the order 
of $TeV$ are expected to be produced within a few years in the 
{\it Large Hadron Collider} being built at CERN. The 
Schwarzschild radii corresponding to these energies also depend on 
the spacetime dimension under consideration, and are related by:
\be
{r_{+(d)}} 
= \le( \f{G_d M}{c^2} \ri)^{\f{1}{(d-3)}}
= \le( \f{ V_{d-4} G_4 M}{c^2} \ri)^{\f{1}{(d-3)}}
= ( V_{d-4} 
{r_{+(4)}} )^{\f{1}{(d-3)}} 
\ee
Once again, using 
$ G_d = {\hbar^{\f{1}{d-3}}c^{5-d}}/{M_{Pl}^{\f{1}{d-2}}} $ 
it can be seen that
although the $r_{+(4)} \approx 10^{-29}~Fm$ is far beyond the realm
of any realistic experiments, $r_{+(d)}\approx 10^{-4}~Fm$ is not. In other words,
the impact parameters can be adjusted to the above value such that the 
colliding protons are within each other's gravitational radii and thus 
form a black hole on colliding. These black holes will then Hawking radiate,
measurements of whose signatures would confirm the existence of higher
dimensions as well as that of black holes. 
 
As shown in \cite{cdm} however, the above set of inferences ought to be 
accompanied by caution. It is widely believed that near the Planck scale,
the Heisenberg Uncertainty Principle (HUP) undergoes modifications and is
replaced by a more refined version known as the Generalised Uncertainty 
principle (GUP) \cite{adler1}: 
\ba
\D x &\geq& \f{\hbar}{\D p}+  (A \l_{Pl})^2 \f{\D p}{\hbar} \\
\D p &=& \f{\hbar\D x}{2 (A \l_{Pl})^2}
~\le[1- \sqrt{1- \f{4 (A \l_{Pl})^2}{\D x^2} } \ri] \la{gup2}
\ea
The constant $A$ is of order unity, but its precise value is theory dependent. 
HUP is recovered in the $\D x/\l_{Pl} \gg 1$ limit. Now, for a Hawking
particle before it is ejected from the horizon has an approximate 
uncertainty of:
$\D x \approx 2r_+~.$ From Eq.(\ref{gup2}), the corresponding uncertainly of 
momentum (which being the only energy scale, is identified with the Hawking 
temperature) 
\ba
T_H &=& { {m^{\f{1}{d-3}}~}{}
~\le[1- \sqrt{1- \f{4 A^2}{m^{\f{2}{d-3}}} } \ri] 
M_{Pl(d)} } 
\la{gup3} \\  
&~\approx& 
\le[ { \f{1 }{m^{1/(d-3)}}}
 + {\f{1}{m^{3/(d-3)}}} + \cdots  \ri] M_{Pl(d)}
\ea
where we have used $r_+ \approx m^{1/(d-3)}$ and $m \equiv {M}/{M_{Pl(d)}}$.
Note that upto overall dimensionless factors, the first term on the right 
represents the usual Hawking temperature, while the second term gives 
leading order corrections to it. The latter being positive, 
the black hole will radiate faster on the brane
according to the Stefan-Boltzmann law \cite{ehm}: 
\be
\f{dm}{dt} 
\propto \mbox{(Area)} \times T_H^4~.
\ee
It can be seen from the exact expression in Eq.(\ref{gup3}) however, that 
the expression inside the square root becomes imaginary and hence
the radiation stops at:
\be
M_{Min} = (4A^2)^{\f{d-3}{2}}~M_{Pl}~.
\ee
Two conclusions follow: first, if the collider energy is below this threshold,
black holes will not form, even if brane-world scenarios are correct.
Second, if the energy is above this threshold, black holes will form and 
radiate, but there will be enormous amounts of missing energies corresponding
to $M_{Min}$. Detection of such missing energies would be a strong 
signature of brane-worlds and black hole remnants.

\subsection{Analog Black Holes} 

Another interesting arena where phenomena analogous to black hole
thermodynamics can be potentially tested is in context of condensed
matter systems, which under suitable circumstances imitate black hole
horizons \cite{unruh,visser}. 
Consider the Navier-Stokes and continuity equations for an 
inviscid and irrotational fluid \cite{visser}:  
\ba
\f{\pa \r}{\pa t } + \nabla \cdot \le(\r v\ri) &=& 0 \\
\r \le( \f{\pa v}{\pa t} + (v \cdot \nabla) v \ri) &=& - \nabla p  \\
\nabla \times v = 0~~\Rightarrow v&=& \nabla\psi ~.
\ea
Perturbation around equilibrium to ${\cal O}(\e)$ is of the form: 
\be
\r = \r_0 + \e\r_1~,~p=p_0 + \e p_1~,~\psi=\psi_0 + \e\psi_1 
~,~\vec v =\vec v_0 + \e \vec v_1
\ee
and combining the first order equation into a single
second order equation yields: 
\ba
&& \f{\pa }{\pa t} \le( c^{-2} \r_0 
\le( \f{\pa\psi_1}{\pa t} + {\vec v}_o \cdot \nabla\psi_1 \ri) \ri)
\nn\\
&=& \nabla \cdot \le( \r_0 \nabla\psi_1 
- c^{-2} \r_0 v_0 
\le( \f{\pa\psi_1}{\pa t} + {\vec v}_0 \cdot \nabla \psi_1 \ri) \ri) 
\la{analog1}
\ea
where 
$c^2 \equiv  \f{\pa p}{\pa \r}$ is the speed of sound in the fluid. 
Defining a $4\times 4$ matrix:
\begin{displaymath}
g^{\m\n}  = \f{1}{\r_0~c}~ 
\left ( \begin{array}{c|c}
-1 & - v_0^j \\
\hline\\
- v_0^i&  (c^2 \d^{ij} - v_0^i v_0^j ) \\
\end{array} \right)
\end{displaymath}
with inverse:
\begin{center}
\begin{displaymath}
g_{\m\n}  = \f{\r_0}{c}~ 
\left ( \begin{array}{c|c}
-(c^2-v_0^2) & - v_0^j \\
\hline\\
- v_0^i&   \d^{ij}  \\
\end{array} \right)
\end{displaymath}
\end{center}
Eq.(\ref{analog1}) can be written as:
\be
\f{1}{\sqrt{-g}} \pa_\m \le( \sqrt{-g} g^{\m\n} \pa_\n \psi_1 \ri) = 0 ~,
\ee
which is just the equation for a scalar field in a curved 
background. In this case, the scalar field represents phonons.  
The analogous infinitesimal line element is:
\be
ds^2= \f{\r_0}{c}~\le[ -(c^2 - v_0^2)~dt^2 
+ d\vec r^2 - 2 \vec v \cdot d\vec r~dt \ri] ~.
\la{analog3}
\ee
Now, choose the following velocity and density profiles:
\be
 v_0 = \sqrt{ \f{2 G_4M}{ r}  }~,~\r_0 = k r^{-3/2}~~
\ee
(where $G_4M$ is a constant) and a new time coordinate:
\be
t' = t 
+ \le[ \f{4G_4 M}{c^3} \arctan \le(\sqrt{\f{2G_4M}{c^2 r}} \ri) 
- 2 \sqrt{\f{2G_4 M r}{c^4}}\ri]~.
\ee 
Then, from (\ref{analog3}): 
\be
ds^2 = \f{k}{c}~r^{-3/2} 
~{\le[ - c^2 
\le( 1 - \f{2G_4M}{c^2 r} \ri) dt'^2 
+ \le( 1 - \f{2G_4M}{c^2 r} \ri)^{-1} dr^2 
+ r^2 d\O_2^2\ri]}  
\ee
which is conformal to the Schwarzschild geometry. The corresponding 
Hawking temperature 
(which is conformally invariant), for typical fluid parameters is 
$T_H = \f{\hbar c^3}{8\p G_4 M} \approx 10^{-4}~K$. 
Similarly, it has been shown that observable {\it superradiance} from these
black hole analogs should result \cite{soumen,gabor2}. It is 
hoped that one would be able to build suitable condensed matter systems in
the future which will demonstrate at least some of the above effects.

\section{Conclusions} 

In this article, we have reviewed various approaches that try to
explain the microscopic origin of black hole thermodynamics. These
include near-horizon conformal field theory,
loop quantum gravity and string theory. While the first two are able to
address realistic Schwarzschild black holes in four dimensions, 
string theory primarily deals with 
extremal RN type black holes. However, the agreements of microscopic and 
macroscopic results pertaining to entropy and Hawking radiation are 
exact and more spectacular in the case of the latter. On the
other hand, whereas CFT and LQG 
pinpoint the location of the degrees of freedom 
in curved spacetime, that give rise to this entropy, 
in string theory it is unclear as to
what the strong coupling counterparts of the $D$-brane 
degrees of freedom are. Moreover, most of string theoretic results pertain to 
five spacetime dimensions. Thus, the results of CFT, LQG and string 
theory appear to be complementary to each other. Although the 
approaches are diverse, since they all attempt to address similar problems, 
it is hoped that 
continuing research in all the fields will someday tell us the exact 
relationship between the degrees of freedom in each approach.

We also studied the problem of information loss for black holes and
some attempts at its resolution. Here too, the resolution is far from
complete. 

Finally, we examined a few experimental scenarios which 
could test the existence of black holes in our universe as well as
imitate black hole thermodynamics in the laboratory. 
We hope that many of the unanswered questions will be 
satisfactorily addressed in the near future.

\vs{.5cm}
\no
{\bf Acknowledgements}

This article is dedicated to the memory of Professor Shyamal Sengupta, who
was an outstanding scientist, rationalist and teacher. 
I thank my past and present collaborators for innumerable discussions, comments
and criticisms over the years, which has helped me in shaping
my understanding of the field. I also  
thank M. Walton for many comments 
which helped in improving an earlier version of the manuscript. 
This work was supported in part by the Natural
Sciences and Engineering Research Council of Canada and funds of The
University of Lethbridge.

\end{document}